\documentstyle[11pt, aas2pp4, psfig]{article}

\def\sp{\hspace{1.5pt}}
\def\amin{\ifmmode^{\prime}\else$^{\prime}$\fi}
\def\asec{\ifmmode^{\prime\prime}\else$^{\prime\prime}$\fi}

\newcommand\xte{{\it RXTE}}

\newcommand\asca{{\it ASCA}}
\newcommand\axaf{{\it AXAF}}

\newcommand\rosat{{\it ROSAT}}
\def\sp{\hskip 1.5pt}
\def\snr{\hbox{N157B}}
\def\psr{\hbox{PSR\sp J0537-6910}}
\def\lmcpsr{\hbox{PSR B0540$-$69}}
\def\wg{Wang \& Gotthelf 1998}
  
\lefthead{Wang \& Gotthelf}
\righthead{The X-ray Pulsar in \snr}

\begin{document}

\title{\large ROSAT HRI Detection of the 16~ms Pulsar PSR J0537-6910 
Inside SNR N157B}

\author{Q. Daniel Wang}
\affil{Dept. of Physics \& Astronomy, Northwestern University}
\affil{2145 Sheridan Road, Evanston,~IL 60208-3112}
\affil{Electronic mail: wqd@nwu.edu}
\affil{and}
\author{E. V. Gotthelf}
\affil{NASA/Goddard Space Flight Center}
\affil{Greenbelt, MD 20771}
\affil{Electronic mail: evg@venus.gsfc.nasa.gov }

\begin{abstract}

Based on a deep ROSAT HRI observation, we have detected a pulsed
signal in the 0.1-2~keV band from \psr\ --- the recently discovered
pulsar associated with the supernova remnant N157B in the Large
Magellanic Cloud.  The measured pulse period 0.01611548182~ms $\pm$
0.02(ns), Epoch MJD 50540.5, gives a revised linear spin-down
rate of $5.1271 \times 10^{-14} {\rm~s~s^{-1}}$, slightly greater than 
the previously derived value.  The narrow pulse shape  (FWHM $\sim 10\%$ duty 
cycle) in the \rosat\ band resembles those seen in both \xte\
and \asca\ data ($\gtrsim 2$~keV), but there is also marginal evidence for
an interpulse. This \rosat\ detection enables us to locate 
the pulsar at R.A., Dec (J2000) = 
$5^h37^m47^s.2, -69^\circ10^\prime23^{\prime\prime}$. With its uncertainty 
$\sim 3^{\prime\prime}$, this position coincides with the
centroid of a compact X-ray source. But the pulsed emission 
accounts for only $\sim 10\%$ of the source luminosity $\sim 2 \times
10^{36} {\rm~ergs^{-1}}$ in the 0.1-2~keV band. These results 
support our previous suggestions: (1) The pulsar is moving at a high 
velocity ($\sim 10^3
{\rm~km~s^{-1}}$); (2) A bow shock, formed around the pulsar, is
responsible for most of the X-ray emission from the source; (3) A
collimated outflow from the bow shock region powers a pulsar wind
nebula that accounts for an elongated
non-thermal radio and X-ray feature to the northwest of the pulsar.

\end{abstract}

\section{Introduction}

PSR J0537-6910 is the most rapidly rotating young pulsar known.  
It was detected serendipitously by Marshall et al. (1998) while searching 
for a pulsed signal from nearby SN
1987A with the {\it Rossi} X-ray Timing Explorer (\xte).  Using archival data 
of the region acquired with the Advanced Satellite for Cosmology and 
Astrophysics (\asca), they further located the pulsed
emission to within the $\sim 1^{\prime}$ radius of the 
supernova remnant (SNR) N157B in the Large Magellanic Cloud (LMC).
This SNR is a moderately
bright X-ray source in the 30 Doradus star formation region and is
clearly imaged with the \rosat\ High Resolution Imager (RHRI) as
isolated diffuse emission (Fig. 1), $\sim 14^{\prime}$ 
east of SN 1987A, and $\sim 15^{\prime}$ northwest of the \lmcpsr.

	Before the discovery of \psr, Wang \& Gotthelf (1998) 
showed that the SNR N157B contains a bright, elongated, and non-thermal X-ray
feature, which dominates the SNR emission and is most likely a pulsar wind 
nebula. Near the southeast end of the feature is a 
compact X-ray source with a spatial extent of $\lesssim 7^{\prime\prime}$.
They suggested that this source represents the pulsar and its interaction 
with the surrounding medium. The predicted SNR age $5\times 10^3$~yrs 
agrees with the characteristic spin-down age of 
\psr\  (Marshall et al. 1998).
These studies have established N157B as the latest addition to the
rare class of Crab-like remnants (e.g., Seward 1989), of which only
three confirmed examples were known previously: the Crab Nebula, the
MSH 15--52 nebula, and LMC SNR B0540$-$693.

\begin{figure}[tbh]
\centerline
{
\hfil\hfil
\psfig{figure=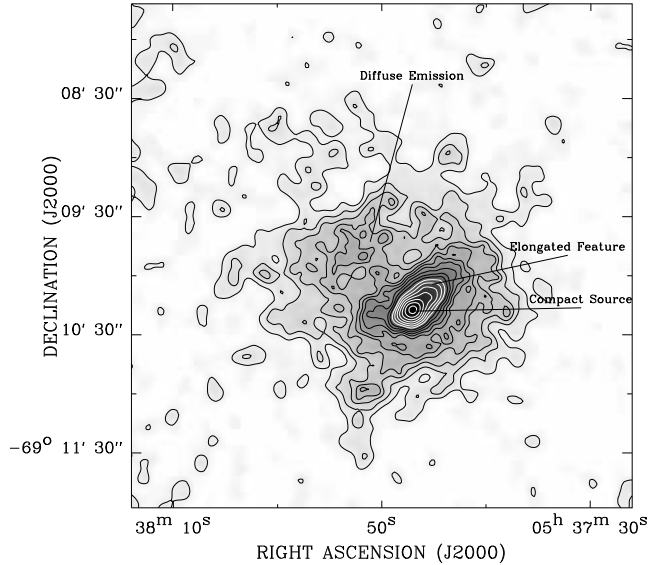,height=3.truein,clip=}
\hfil\hfil
}
\caption{\protect\footnotesize Coadded RHRI image of \snr. The image
is adaptively smoothed with a Gaussian of adjustable size to achieve 
a constant local count-to-noise ratio of $\sim$ 6 over the field. Each
contour is 2$\sigma$ above its {previous} level. The lowest level is
about 2$\sigma$ above the local background of 
$\sim 7 \times 10^{-3}$ counts s$^{-1}$ arcmin$^{-2}$, while the highest 
one is at 1.06 counts s$^{-1}$ arcmin$^{-2}$.
}
\end{figure}

	We report in this Letter the detection of pulsed soft X-ray
emission from \psr, using a deep RHRI observation recently acquired.
This detection is the first in the energy band below
2~keV. This band-pass is not available to X-ray detectors with poor
spatial resolution, because of the overwhelming soft X-ray emission from both 
the pulsar
wind nebula and the diffuse thermal hot gas of N157B. Pulse phase-resolved
high resolution imaging capability of the RHRI further enables us to locate 
the pulsar with a linear position accuracy of $\lesssim
3^{\prime\prime}$, a factor of $> 10$ better than the previous
measurements. The pulsar location confirms our early suggestion of the 
compact X-ray source as the site of the pulsed emission.
The accuracy of the location may also allow for identification of 
pulsar counterparts in other wavelength bands.

\section{Observation and Data Reduction}

	The new RHRI observation was taken as part of our ongoing X-ray 
investigation of the 30 Dor region. The high resolution 
X-ray image of N157B as shown in Fig. 1 is a co-add of the data from 
\rosat\ observation Sequence No. rh600228n00 (30~ks exposure taken 
between Dec. 12, 1992 - Jun. 14, 1993; Wang 1995; Wang \& Gotthelf
1998), rh400779n00 (26 ks between Aug. 21, 1996 - Jan. 27, 1997) and 
rh400779a01 (79~ks between April 3-29, 1997). The first two exposures 
are, however, too short to be effective for detecting the pulsar. Our timing 
analysis was thus based exclusively on the latest observation. During the 
observation, the satellite was wobbled, and \psr\ was located very close ($\sim
2\farcm7$) to the time-averaged optical axis. The expected Point Spread 
Function (PSF) is about 6$^{\prime\prime}$ (FWHM).

	We processed the data to optimize their timing and positioning
accuracy. First, we deleted time intervals 
with unstable detector roll angles and/or with fewer than 3 guide stars
for position tracking, and shifted counts in individual observing segments 
(typically about 1~ks long) to have a common centroid position 
of the compact source (Fig. 1). This process improved the photon position 
accuracy by about 1$^{\prime\prime}$ in terms of the FWHM of the PSF. 
The source is not strong enough, however, to allow 
for a wobble phase-resolved centroid correction; thus significant wobble 
phase-dependent pointing errors may still be present in the data.
Second, we corrected the absolute astrometry 
of the data by matching the X-ray and optical positions of three identified 
objects (the foreground stars cal 69 and cal 71 as well as the LMC 50 ms 
pulsar B0540-69; Wang 1995). After this correction, the residual differences 
between the X-ray and optical positions of these three objects are all less 
than $\sim 0\farcs4$. Third, to reduce the non-cosmic contamination
we selected data only within the pulse height channel 2-10. Fourth, we 
converted the arrival time of each photon to the solar system barycenter 
using the JPL DE200 ephemeris.

\section{Data Analysis and Results}

	For the timing analysis, we extracted counts within a circle of 
7$^{\prime\prime}$ radius around the centroid of the compact source --- 
the suspected location of the pulsar (Wang \& Gotthelf 1998). These counts 
were to be used to search for the periodic signal around the pulse 
frequency ($\nu$) extrapolated from the previously measured values and the 
mean frequency derivative $\dot{\nu} = -1.9746 \times 10^{-10} {\rm~s^{-2}}$ 
(Marshall et al. 1998). However, we found that a small uncertainty 
in $\dot{\nu}$ could cause a significant phase shift over the observation's 
total duration of 2234676~s, during which individual observing segments were 
separated by various data gaps resulting from the satellite passage through 
the South Atlantic Anomaly,  Earth occultation, etc. Based on the scatter of 
measured pulse frequencies around the linear spindown relation 
(Marshall et al. 1998), we estimated that the $\dot{\nu}$
uncertainty is $\sim 0.04\%$, which could cause a phase shift of 
0.2 cycle. Thus we removed several small observing segments (total 266 counts)
that were largely separated from the main middle portion of the observation. 
The remaining 1288 counts are all within MJD 50545.0-50554.8 (849270~s). 
Over this time interval the phase shift of 0.03 cycle is sufficiently small, 
compared to the pulse width of $\sim 0.1$ cycle (FWHM). 

	We then constructed a frequency-gram 
($\chi^2$ vs. $\nu$) of the selected counts to search for the periodic 
signal. For each trial frequency, we 
folded the counts into 10 phase bins to optimize the signal in the counting
noise-dominated data and calculated the $\chi^2$ statistic of the resultant 
profile to test against the null hypothesis of no signal.
A significant signal with $\chi^2 \sim 60$ ($\gtrsim 12\sigma$) is clearly 
present in the frequency-gram. Due to the presence of noise in 
the data, the peak with the maximum $\chi^2$ value, however, is not 
necessarily the optimal measurement of the signal period. 
In fact, the exact peak period depends slightly on the frequency step used 
to construct the frequency-gram. Both the noise and the frequency step 
dependency can be suppressed by convolving the frequency-gram with an 
appropriate filter.
We generated such a filter directly with the smoothed signal peak 
in the frequency-gram with the smallest frequency step. 
Because the shape of the peak depends primarily
on the pulse profile and on the total observing duration, the filter can 
be re-binned for frequency-grams with larger
steps. We adopted the signal peak {\sl centroid} of the filtered 
frequency-gram as our period measurement (Table 1), which is nearly
 independent of its chosen frequency step and is insensitive to the exact 
shape of the filter. An analog of this approach is the position centroid 
determination of a source in a 2-D X-ray image convolved with its PSF, 
or simply with a Gaussian filter.

	The uncertainty in our period measurement of the pulsar is determined 
through Monte Carlo simulations, based on the 
measured pulse profile before the binning and on the actual observing 
segments. Each of our 1000 simulated data sets had the same 
number of counts as in the observed data, and were analyzed in the exactly 
same way. The dispersion in the obtained  
period distribution was adopted as an estimate of the uncertainty in our 
measurement of the pulsar period (Table 1).

\begin{table}[bth]
\begin{center}
\begin{tabular}{lr}
\multicolumn{2}{c}{\bf Table 1: Pulsar Parameters}
\\
\\
R.A. (J2000)	&$5^h37^m47\fs2$\\
Dec. (J2000)	&$-69^\circ10^\prime23^{\prime\prime}$\\
Frequency $\nu$ (MJD 50540.5) &$62.05213169(\pm 8){\rm~s^{-1}}$\\
Mean $\dot{\nu}$ (1993-1997) &$-1.97493 \times 10^{-10} {\rm~s^{-2}}$\\
\ or Mean $\dot P$ &$5.1271 \times 10^{-14} {\rm~s~s^{-1}}$\\
Pulse Width (FWHM)&$\sim 2$ ms \\
RHRI Count Rate &$3.0 \times 10^{-3} {\rm~counts~s^{-1}}$\\
Pulsed Flux (0.1-2~keV)$^a$&$\sim 6 \times 10^{-13}$\\
& ${\rm~ergs~cm^{-2}~s^{-1}}$\\
\end{tabular}
\tablenotetext{a} {Assuming a power law spectrum of energy slope 
$\alpha = 0.6$ and an absorption of $1 \times 10^{22} {\rm~cm^{-2}}$  
(40\% solar abundance; Marshall et al. 1998.; \wg).}
\end{center}
\end{table}

\begin{figure}[tbh]
\centerline{
\psfig{figure=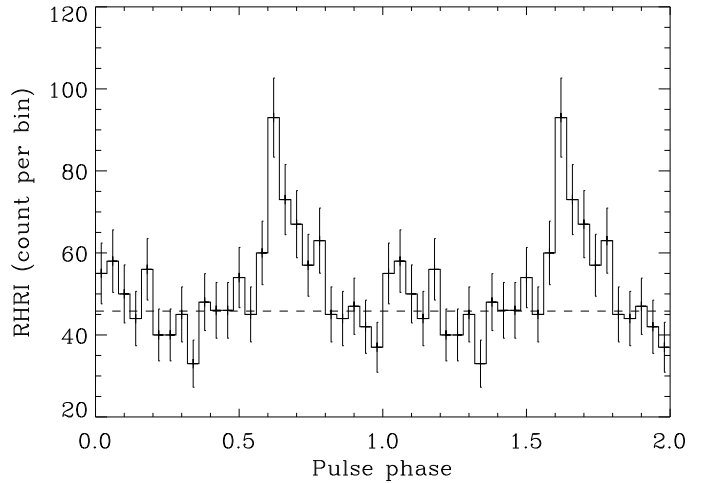,height=2.8truein,angle=0.0,clip=}
}
\caption{\protect\footnotesize Pulse profile of \psr\ in
the 0.1-2~keV band from the RHRI observation. The dashed line refers to 
the mean number of counts per bin averaged 
outside the 0.55-0.8 on-pulse phase interval. For a better illustration of 
the profile in this plot, a full cycle is duplicated and 25 bins per cycle 
are used. 
}
\end{figure}

	Fig. 2 shows the pulse profile. 
The pulse profile is narrow (FWHM $\sim 0.1$ duty cycle) and 
resembles the ones seen in the previous detections (\S 1), 
all of which are in energy bands $\gtrsim 2$~keV. The pulsed 
intensity is about 30\% of the total intensity during the full-width duty 
cycle of $\sim 0.25$, or about 10\% if averaged over the full cycle. 
Fig. 2 also indicates the presence of a possible weak interpulse, about 0.5 
phase-shifted from the main pulse; but the significance of the detection 
is low ($\sim 3 \sigma$).

\begin{figure}[bth]
\centerline{
\psfig{figure=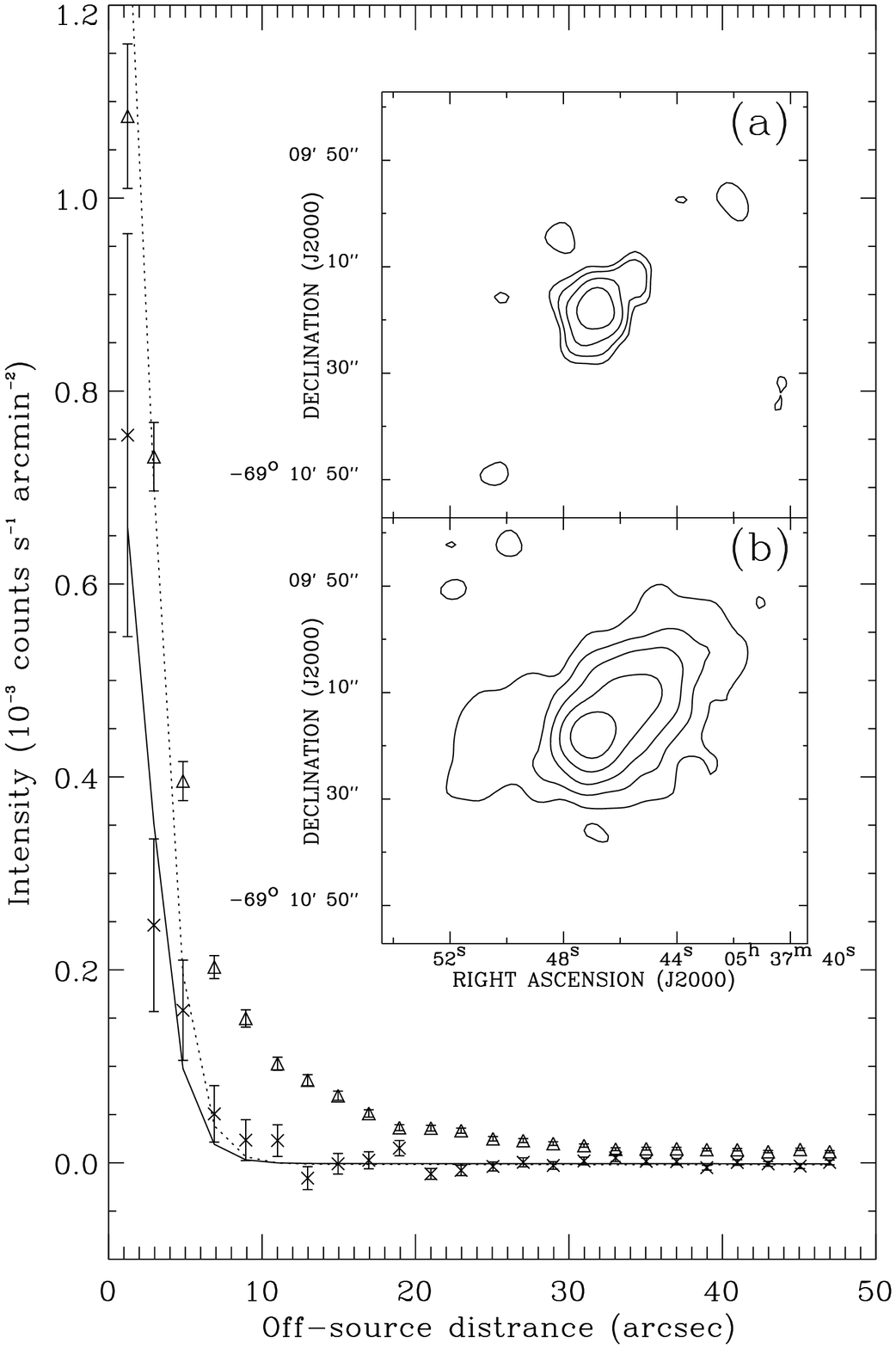,height=5.5truein,angle=0.0,clip=}
}
\caption{\protect\footnotesize The pulsed and unpulsed soft X-ray emissions 
from N157B. The pulsed emission was derived from the RHRI data within the 
full-width of 0.25 duty cycle and after subtracting normalized off-pulsed
data. The average pulsed and unpulsed radial intensity profiles are marked 
with {\sl crosses} and {\sl triangles}, respectively. The pulsed emission 
data are 
reasonably well-fitted ($\chi^2/d.o.f. = 24.4./22$) with the PSF shown as the 
solid curve. The slight excess of the data points above the PSF
between 5-10$^{\prime\prime}$ can be explained by telescope wobble-related
aspect errors (\S 2). For comparison, the dotted curve shows the PSF 
normalized to the intensity of the unpulsed intensity profile. The inserted 
plot shows the intensity images of pulsed (a) and unpulsed (b) emissions. 
The contours are at 3, 6, 12, 24, and 48 $\times 10^{-2} 
{\rm~counts~s^{-1}~arcmin^{-2}}$.
}
\end{figure}

	There is little doubt that the pulsar is indeed within our adopted 
count collecting aperture. We find that the pulsed signal diminishes with the 
increase of the aperture radius, apparently due to the proportional increase 
of unpulsed emission from the remnant. 
To determine the pulsar position accurately, we produced an
image corresponding to the pulsed emission only. The insert of Fig. 3 
includes an on-pulse intensity image with the off-pulse data subtracted. The
intensity was calculated within the phase interval between 0.55 and 0.80.
The average radial intensity distribution around the source centroid 
in the image is consistent 
with the PSF (Fig. 3). We thus adopt the source centroid as the position of the
pulsar.  The statistical uncertainty in the position is 
$\sim 1^{\prime\prime}$. But we could not rule out a 
systematic error up to $\sim 2^{\prime\prime}$, introduced by such effects as
the asymmetric shape of off-axis PSF, which may affect the
accuracy of our astrometry correction (\S 2). Fig. 3 also includes an 
off-pulsed intensity image which clearly shows a strong intensity peak, 
probably barely resolved, and an elongated feature that extends at
least $\sim 30^{\prime\prime}$ to the northwest.

\section{Discussions and Conclusions}

	The above results, as summarized in Table 1, provide an 
accurate position 
of  \psr\ inside the SNR N157B, represent the first detection of 
the pulsar in the 0.1-2~keV band, and extend the period measurement of 
the pulsar to Epoch 1997. 
This latest measurement, together with the earliest
\asca\ observation (June 13, 1993; Marshall et al. 1998), gives an average 
period derivative (Table 1) that is slightly
greater than $5.126 \times 10^{-14} {\rm~s~s^{-1}}$ as derived from the \asca\ 
and 1996 \xte\ observations (Marshall et al. 1998). This discrepancy is 
consistent with the presence of a giant pulsar glitch ($\delta \nu/\nu \sim
10^{-6}$) occurred between the latter pair of observations. The pulsed 
emission accounts for only $\sim 3\%$ of the total luminosity of SNR N157B
in the RHRI 0.1-2~keV band (Wang \& Gotthelf 1998). This pulsed contribution
is substantially smaller than  $\sim 10\%$ in the \asca\ ($2-10$~keV) band,
but is consistent with the extrapolated 0.1-2~keV flux from the high energy
band (Table 1). The pulsed emission spectrum (power law energy slope
$\alpha \approx 0.6$) is significantly harder than the mean remnant spectrum 
($\alpha \approx 1.5$; \wg). The 0.1-10~keV luminosity of the pulsed
emission (into 4$\pi$) is $\sim 3.3 \times 10^{35} {\rm~ergs~s^{-1}}$, or 
$\sim 4 \times 10^{-4}$ of the pulsar's spin-down luminosity.
 
	These new results are fully consistent with our early
suggestion about the dynamic relationship between the pulsar and the Crab-like 
remnant N157B (Wang \& Gotthelf 1998). First, as 
indicated by a separation of $\sim 12^{\prime\prime}$ between the X-ray 
and radio emission peaks of the remnant, the pulsar is likely moving at 
a high velocity ($\sim 10^3 {\rm~km~s^{-1}}$). Second, this motion has led to
the formation of a bow shock around the pulsar. The enhanced synchrotron
radiation of relativistic particles in the shocked pulsar wind accounts for the
unpulsed component of the compact source. Third, a collimated outflow, 
driven by the pressure difference between the bow shock and the 
ambient medium (Wang \& Gotthelf 1998; see also Wang et al. 1993), powers
an extended pulsar wind nebula where most of the pulsar spindown energy has 
been dumped. The separation between the pulsar and the main pulsar wind
nebula explains the elongated feature seen in both radio and X-ray. While 
both the pulsar position and the small pulsed contribution to the compact 
source are confirmed here, we continue to study the relationship between
\psr\ and N157B using archival data in various
energy bands and with upcoming \axaf\ observations.

\begin{acknowledgements}
{\noindent \bf Acknowledgments} --- We thank J. Halpern for suggesting
the use of IRAF/PROS for the barycenter correction and J. Finley, 
the referee, for communicating his comments directly to us.
Q.D.W. is supported partly by NASA LTSA grant
NAG5-6413. E.V.G. is supported by USRA under NASA contract NAS5-32490.
\end{acknowledgements}

\end{document}